\documentclass{sig-alternate}
\usepackage{mathptmx} %

\usepackage{fancyhdr}
\usepackage[normalem]{ulem}
\usepackage[hyphens]{url}
\usepackage[sort,nocompress]{cite}
\usepackage[final]{microtype}
\usepackage[keeplastbox]{flushend}
\usepackage{makecell}
\usepackage[bookmarks=true,breaklinks=true,colorlinks,linkcolor=black,citecolor=blue,urlcolor=black]{hyperref}
\usepackage{multirow}
\usepackage{tikz}
\usepackage{pgfplots}
\usepackage{comment}

\fancypagestyle{firstpage}{
    \fancyhf{}
    
    \fancyhead[C]{}
    \fancyfoot[C]{}
}%

\usepackage{tikz}
\newcommand*\circled[1]{\tikz[baseline=(char.base)]{
            \node[shape=circle,draw, fill=black, text=white, inner sep=0.1mm] (char) {#1};}}

\usepackage[ruled,vlined]{algorithm2e}

\usepackage{tikz}
\def\checkmark{\tikz\fill[scale=0.4](0,.35) -- (.25,0) -- (1,.7) -- (.25,.15) -- cycle;}

\newcommand{\ignore}[1]{}

\usepackage{amsmath}

\usepackage{comment}
\usepackage{booktabs}
\usepackage{todonotes}

\newcommand{\name}{PARADISE}

\newcommand{\basic}{basic}
\newcommand{\educated}{educated}
\newcommand{\advanced}{advanced}
\newcommand{\Basic}{Basic}
\newcommand{\Educated}{Educated}
\newcommand{\Advanced}{Advanced}

\newcommand{\rocket}{io-baseline}
\newcommand{\baseline}{ooo-baseline}
\newcommand{\secure}{PARADISE}
\newcommand{\aggressive}{random-aggressive}
\newcommand{\isoperf}{random-iso-perf}
\newcommand{\isosecu}{random-iso-security}

\newcommand{\isoperb}{$12 \times$}
\newcommand{\isopere}{$12 \times$}
\newcommand{\isopera}{$5.5 \times$}

\newcommand{\isoseca}{$34 \times$}
\newcommand{\aggb}{$122 \times$}
\newcommand{\agge}{$11 \times$}
\newcommand{\agga}{$38 \times$}
\newcommand{\parab}{$261 \times$}
\newcommand{\parae}{$12 \times$}
\newcommand{\paraa}{$34 \times$}

\title{Mitigating Power Attacks through Fine-Grained \\ Instruction Reordering }%

\numberofauthors{6}
\author{
\alignauthor
Yun Chen\titlenote{} \\
    \affaddr{National University of Singapore} \global\titlenotecount 0\relax %
\alignauthor
Ali Hajiabadi\titlenote{Yun Chen and Ali Hajiabadi contributed equally to this work.} \\
    \affaddr{National University of Singapore}
\alignauthor
Romain Poussier \\
    \affaddr{Nanyang Technological University}
\and
\alignauthor
Andreas Diavastos \\
    \affaddr{Universitat Politècnica de Catalunya}
\alignauthor
Shivam Bhasin \\
    \affaddr{Temasek Laboratories, Nanyang Technological University}
\alignauthor
Trevor E. Carlson \\
    \affaddr{National University of Singapore}
}
\begin{document}

\maketitle
\thispagestyle{firstpage}
\pagestyle{plain}

\begin{abstract}

Side-channel attacks are a security exploit that take advantage of information leakage. They use measurement and analysis of physical parameters to reverse engineer and extract secrets from a system. Power analysis attacks in particular, collect a set of power traces from a computing device and use statistical techniques to correlate this information with the attacked application data and source code. Countermeasures like just-in-time compilation, random code injection and instruction descheduling obfuscate the execution of instructions to reduce the security risk. Unfortunately, due to the randomness and excess instructions executed by these solutions, they introduce large overheads in performance, power and area.

In this work we propose a scheduling algorithm that dynamically reorders instructions in an out-of-order processor to provide obfuscated execution and mitigate power analysis attacks with little-to-no effect on the performance, power or area of the processor. We exploit the time between operand availability of critical instructions (\textit{slack}) to create high-performance random schedules without requiring additional instructions or static prescheduling. Further, we perform an extended security analysis using different attacks. We highlight the dangers of using incorrect adversarial assumptions, which can often lead to a false sense of security. In that regard, our advanced security metric demonstrates improvements of \paraa{}, while our basic security evaluation shows results up to \parab{}. Moreover, our system achieves performance within 96\% on average, of the baseline unprotected processor.

\end{abstract}

\section{Introduction}

Timing, electromagnetic (EM) and power analysis attacks, also called side-channel attacks, exploit physical parameters of the processor to extract secret information from a running program in the system. Power analysis attacks are the most common as they are relatively easy and cheap to execute as they do not require any special equipment or knowledge of the internal design of the system. Power analysis attacks take advantage of the synchronization and high correlation of the power consumption with the instructions and data being processed to identify patterns in the victim's program behavior. These patterns are then analyzed to reverse engineer the program execution and extract secret information~\cite{svf}.

Many countermeasures have been proposed to combat power analysis attacks in all levels of the design stack~\cite{savat, kar2017improved, racoon, rijid, erist,ghostrider, random-code-inject, ansi,falsekey}. We categorize these countermeasures into two generic mitigation techniques: (1) \emph{hiding} and (2) \emph{masking}. \emph{Hiding} countermeasures limit or hide the information available to an adversary. In other words, \emph{hiding} lowers the available signal-to-noise ratio (SNR) to an adversary. SNR suppression can be achieved by balancing or suppression techniques where relations between power consumed are weakened for executed data or instruction (e.g. power balancing, physical signal suppression, noise amplification) or by randomizing the execution sequence (e.g. jitter, randomization of execution order).
Thus, \emph{hiding} is a system-wide solution that alters the power consumption of the system in a way that hides the actual power consumed by the instructions and the data being processed.

\emph{Masking}, on the other hand, splits any sensitive intermediate variable into several statistically independent shares, similar to the principles of Shamir's secret sharing\cite{shamir1979share}. An adversary can learn nothing about the sensitive intermediate variable, unless all the shares are available. As the shares are processed independently, \emph{masking} effectively removes all leakage dependencies for the lowest statistical moments, which increases the attack complexity. Often, design constraints must be put in place to guarantee independence of shares in data processing.
The processing of independent shares requires additional computation, leading to a non-negligible overheads. While the \emph{masking} technique can provide strong guarantees from a cryptographic point of view, it is effective only in the presence of noise. Thus \emph{masking} and \emph{hiding} are complementary countermeasures, where hiding provides the ideal low SNR environment for masking to be effective.

In this work, we focus on solutions and methodologies to improve secret \textit{hiding}. In fact, the current solutions that implement the \emph{hiding} technique make significant sacrifices in a number of important areas. Some significantly increase the performance, power or area of the system to mitigate an attack~\cite{hwang2006aes, kar2017improved, ansi, althoff2018hiding, param, ge2020power, rijid}. The overheads in performance and power reach up to 300\%, while the area can be twice the size of the original baseline core. Other solutions either lack generality~\cite{falsekey} or require compiler support~\cite{shuffler}. In general, we observe that no one single solution offers a competitive package that provides high-levels of security with minimal impact on the performance, power, design cost or compatibility of the system. 

With this work, we aim to overcome these limitations of previous hiding techniques. We introduce a new countermeasure that leverages the scheduling information of dynamic instructions in an out-of-order processor to break the correlation to physical parameters with minimal overhead (3.7\% on performance, 1.1\% on power and 0.7\% on area). To implement our proposed hiding technique in an efficient way, we propose an instruction scheduling technique that monitors the execution of an application and intelligently randomizes the issue of instructions to produce different instruction schedules in every loop without affecting the performance. More specifically, we add a low-cost structure (\textit{Slack Unit}) in the processor pipeline that records the time difference between an instruction's operand production (slack) and uses it to inject a random artificial delay (that is no larger than this original slack) to randomize the issue of non-critical instructions. The result is a low-overhead (both in energy and area, as well as performance), general-purpose and hardware-only technique that obfuscates instruction execution to break the power consumption correlation that is normally observed by the adversary.

In this paper we make the following contributions:
\begin{itemize}
    \item A general-purpose out-of-order core design that provides a significant security improvement (\paraa{} with an advanced adversary, and \parab{} when the attacker is using basic methods) against power analysis attacks to all executing applications with minimal power and area overheads (1.1\% and 0.7\% respectively); 
    
    \item A dynamic instruction scheduling technique that intelligently combines instruction slack with randomness to avoid significant performance degradation (less than 3.7\%), while strengthening the security of the system;
    
    \item An advanced, comprehensive security evaluation of the proposed solution. We present three different security evaluation techniques in an attempt to raise the bar for architectural security evaluation.
\end{itemize}

This paper is structured with an overview in Section~\ref{sec:motivation} and describe the proposed architecture in Section~\ref{sec:microprocessor}. In Section~\ref{sec:security} we outline our proposed evaluation methodology and in Section~\ref{sec:evaluation} we present our findings. In Section~\ref{sec:related} we outline the state-of-the-art and conclude with Section~\ref{sec:conclusions}.

\section{Motivation and Overview}
\label{sec:motivation}

Power analysis attacks exploit the synchronization of instruction execution and its correlation with power consumption to uncover secret information from a running application. A power analysis attack is performed by collecting a set of power traces from a processor during victim application execution. The number of power traces required to reveal this secret information depends on how secure the processor is, where regular patterns can more easily be used to extract this data. Using statistical analysis, the adversary correlates the power consumption with the executed instructions and the data being processed to uncover the secret information of the application. Such attacks are effective due to the deterministic behavior of the processing units, and are possible because the adversary can reason about the processor's behavior with a relatively small set of power measurements. As we show later, and as demonstrated by previous works~\cite{mcfarlinspeculationdynamism2013}, even modern general-purpose out-of-order processors demonstrate a high level of regularity, making them relatively easy to attack.

A common countermeasure for these types of attacks is to execute instructions in an unpredictable way that makes it difficult for the adversary to correlate the observed power with the code being executed. Many solutions have been proposed in this direction~\cite{hwang2006aes, kar2017improved, ansi, althoff2018hiding, param, ge2020power, rijid, shuffler, falsekey},
however, most of them are application-specific, target simple dedicated-to-encryption hardware, or do not apply generically to processor platforms. Such solutions also tend to result in large performance and/or power overheads or require significantly more hardware to implement. One of the biggest challenges in designing secure processors is finding efficient generic solutions that can protect all applications without affecting application performance.

In this work, we focus on the general case and target high performance out-of-order general-purpose processors. To this end, we identify three major requirements needed to successfully design both a secure and efficient general-purpose processor: (1) desynchronize the instruction execution (create non-deterministic behavior) to increase the noise in the collected power traces without affecting the performance, (2) design a generic solution that allows for secure execution of all running applications, and (3) implement an efficient design that demonstrates minimal power and area overheads.

\begin{figure}[t]
    \centering
    \includegraphics[width=0.35\linewidth]{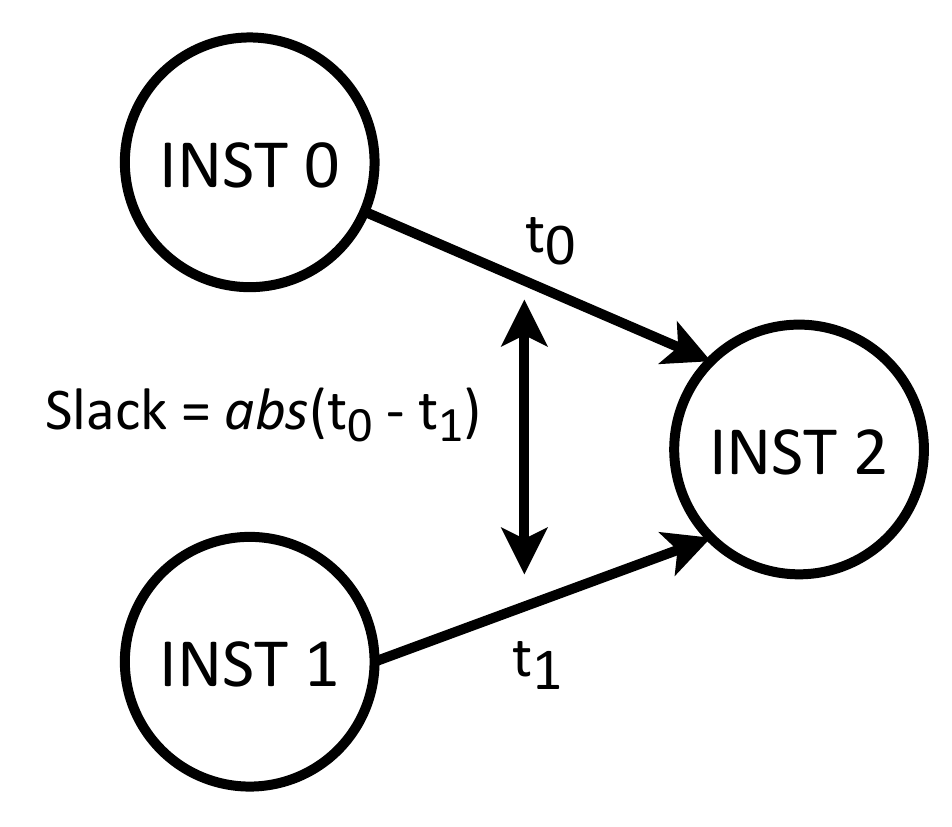}
    \caption{Computing slack for instructions. The slack for INST 2 is the absolute difference between the time that INST 0 and INST 1 produce their result.}
    \label{fig:slack}
\end{figure}

The key insight of this paper is that to efficiently offer security in a general-purpose processor without affecting its performance we must \textit{randomize the ordering of instructions that are off the critical path of the execution}. As explained earlier, increase in security can result from randomizing the instruction execution order. In this work, we introduce randomization by delaying instruction issue time. Doing so desynchronizes the execution, enables non-deterministic behavior, and therefore increases the noise as seen in the power measurements. Consequently, the added noise makes it harder for the adversary to find the highly distinguishable statistical dependencies that will uncover secret information. To maintain baseline performance however, reordering must be done only on non-critical instructions. In this work, we define criticality by the time the input operands of an instruction are produced. In the example of Figure~\ref{fig:slack}, instruction INST 2 has two input operands that are produced at different times. Assuming that INST 0 produces its output earlier than INST 1 then, in this pair of instructions, INST 0 is non-critical because delaying its execution until INST 1 produces its output will not change the starting time of INST 2. Delaying INST 1 however, will delay the execution of INST 2 therefore, we consider INST 1 a critical instruction.%

Our proposed solution injects a delay only to non-critical instructions and within certain timing boundaries that will minimize performance overheads. We call this time margin \textit{slack}. The slack is defined as the time difference between the generation of an instruction's operands. In the example of Figure~\ref{fig:slack}, INST 0 and INST 1 produce their data for INST 2 at time $t_{0}$ and $t_{1}$ respectively. The absolute difference of the two times ($abs(t_{0}-t_{1})$) is the slack for these two instructions. If we assume that $t_{1} > t_{0}$, then INST 0 is considered to be non-critical for the execution of INST 2 and this information will be saved for the next time we execute this instruction. At that point, a random delay, not larger than the slack, will be introduced to increase the issue time of INST 0. This in turn increases instruction desynchronization, making it more difficult for an adversary to extract secret information.

Note that, although throughout this paper we refer to power analysis attacks only, the proposed technique is also applicable to other side channels that use similar attack methodologies like EM and timing attacks.

\begin{figure}[t!]
    \centering
    \includegraphics[width=1\linewidth]{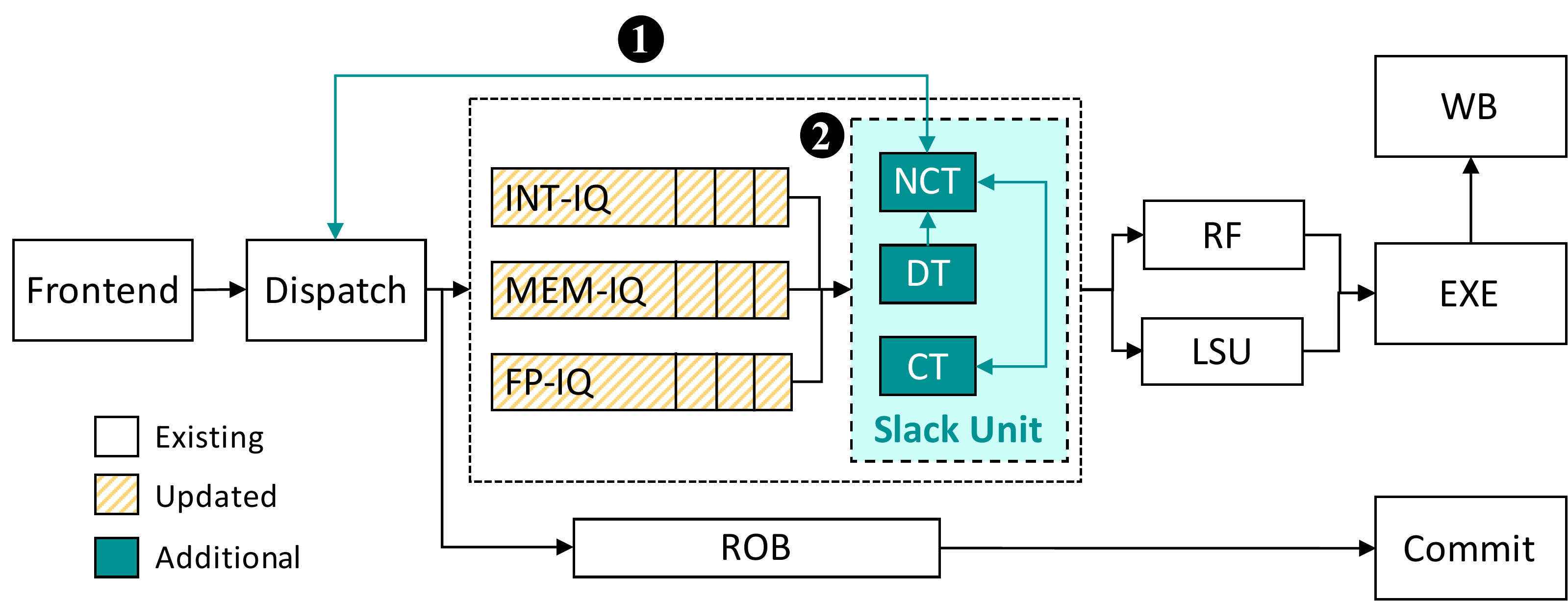}
    \caption{The microarchitecture of \name{}.}
    \label{fig:ppl}
\end{figure}

\section{\name{} Microarchitecture}
\label{sec:microprocessor}

In this section, we present an efficient general-purpose processor that detects the slack and randomly delays non-critical instructions to desynchronize the execution without affecting the critical path and successfully maintaining high performance. Non-critical instructions are delayed before being issued to the execution units to randomize their execution and their access to the register file or memory. To achieve this, we propose \textit{\name{}}, an out-of-order processor that implements a novel secure instruction scheduling policy. %

Figure~\ref{fig:ppl} shows the microarchitecture of \name{}. It is built on top of SonicBoom~\cite{zhaosonicboom}, a RISC-V out-of-order core with a 7-stage pipeline. \name{} implements a new structure called the Slack Unit that dynamically records the slack of instructions and communicates the appropriate delay to be injected to selected non-critical instructions. The Issue Queues (IQs) and the out-of-order scheduler are updated to communicate instruction information to the Slack Unit and delay the issuing of instructions that were previously injected with a delay respectively. These additions don't affect the pipeline stages of the processor as they are parallel operations that execute seamlessly with the rest of the processor operations. Hence, \name{} retains the frequency of the baseline SonicBoom core.

A dispatched instruction queries the Slack Unit, using its PC, and in case of a hit a delay, is injected to the instruction and the scheduler is notified that the current instruction will be issued with a delay (step~\circled{1}). The delay is applied only after all its operands are produced. When an instruction is issued for execution, the Slack Unit stores its producers and the new slack between them (step~\circled{2}).

\subsection{Slack Unit}

\begin{figure}[t]
    \centering
    \includegraphics[width=\linewidth]{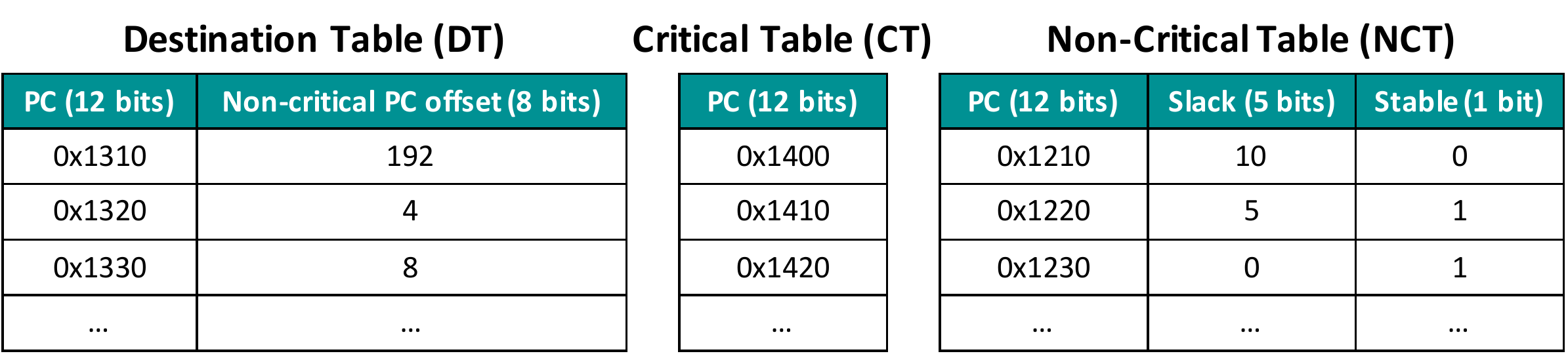}
    \caption{The Non-Critical Table (NCT) stores non-critical instructions and their slack, the Critical Table (CT) stores critical instructions and the Destination Table (DT) stores the currently dispatched instructions.}
    \label{fig:sram}
\end{figure}

The Slack Unit is built using three set-associative memory structures to store different runtime information of the program, as shown in Figure~\ref{fig:sram}. All structures use a Least Recently Used (LRU) replacement policy. The structures are:

\begin{itemize}

    \item \textbf{Destination Table (DT)}: Holds the instructions that are being issued and their non-critical producer. The DT is queried by issued instructions to check whether their non-critical producers are the same in different appearances;
    
    \item \textbf{Non-Critical Table (NCT)}: Holds the non-critical instructions and their corresponding slack. For performance reasons, we also use these entries to indicate whether the instruction has been consistently a non-critical instruction (stable). If so, we use the stored slack as an upper bound for selecting a random delay for the corresponding instruction;
    
    \item \textbf{Critical Table (CT)}: Holds a list of instructions that were marked as critical. This is necessary in order to handle criticality conflicts in cases where an instruction is a critical producer for an instruction but a non-critical producer for another.
    
\end{itemize}

In our experiments with several encryption applications (AES-128 engine, SHA3, RSA, etc.), we found that an 8-bit offset is sufficient to represent the PCs of non-critical instructions in the DT because the PCs of instructions are not far from their dependent instruction PCs.

\subsection{Criticality and Slack Detection}

As explained in Section~\ref{sec:motivation}, slack is defined as the time difference between the production of an instruction's operands.

Assuming the same example of Figure~\ref{fig:slack}, the slack of instruction INST 2 will be the absolute difference of the times that its producers (INST 0 and INST 1) will return their result, represented by the Equation~\ref{eq:slk}.

\begin{equation}
\label{eq:slk}
    slack(INST\;2) = abs(t_0 - t_1)
\end{equation}

When a new instruction is issued, an entry is allocated for it in the Destination Table (DT). When the same instruction completes its execution, the Slack Unit detects the criticality it produces and calculates the slack. The critical instruction is stored immediately in the Critical Table (CT). At the same time the Non-Critical Table (NCT) is checked and if the current critical instruction matches with a previous NCT entry, the entry is dropped as the critical status supersedes. Before storing the non-critical instruction we must first query the CT for possible criticality conflicts. A criticality conflict occurs when an instruction is a producer for two or more instructions and its criticality status is different for each one of them. In this case the producing instruction is not stored in the NCT. In the absence of a conflict with the CT, the producer will be stored in the NCT table coupled with the calculated slack and marked as not stable. If the instruction was already stored in the NCT and the new slack is smaller than the previously stored one, we update the entry and add the new (smaller) slack. When an issued instruction hits the DT (indicating that it appeared before), we verify that the criticality status of its producers is the same and update the non-critical status in the NCT to stable. If their criticality status changed, we make the appropriate changes in all Slack Unit structures. %

Although false hits in the Slack Unit can happen when context-switching processes with the same PCs, solutions exist (flushing the Slack Unit on context switches, or tagging tables with process-specific information), but the evaluation of these solutions fall outside the scope of this work.

\subsection{Delay Injection}
\label{sec:iq}

To find whether a newly dispatched instruction is to be reordered, the Slack Unit is queried (specifically the NCT) with its PC and in case of a hit, the appropriate delay is returned and injected to its issue slot. This delay is only applied just before the instruction is marked by the out-of-order scheduler as ready to execute. That is, after all its input operands have been produced. As soon as its operands are produced, the instruction will be marked to wait for another \textit{delay} number of cycles. If the delay is marked in NCT as \texttt{stable}, it will be used as an upper bound to select a random value between 0 and the stable delay. If the delay is not stable, we label this delay as being in the the \textit{unstable phase}, and we don't randomize this delay. This is done to add an additional layer of randomization in the desynchronization of the execution. However, the delay will still be used as is, without randomization, if the slack is unstable to avoid adding unnecessary performance overheads in the execution. Note that the Slack Unit is constantly learning to determine the stable delay.

\subsection{Example}

\begin{figure}[t]
    \centering
    \includegraphics[width=0.75\linewidth]{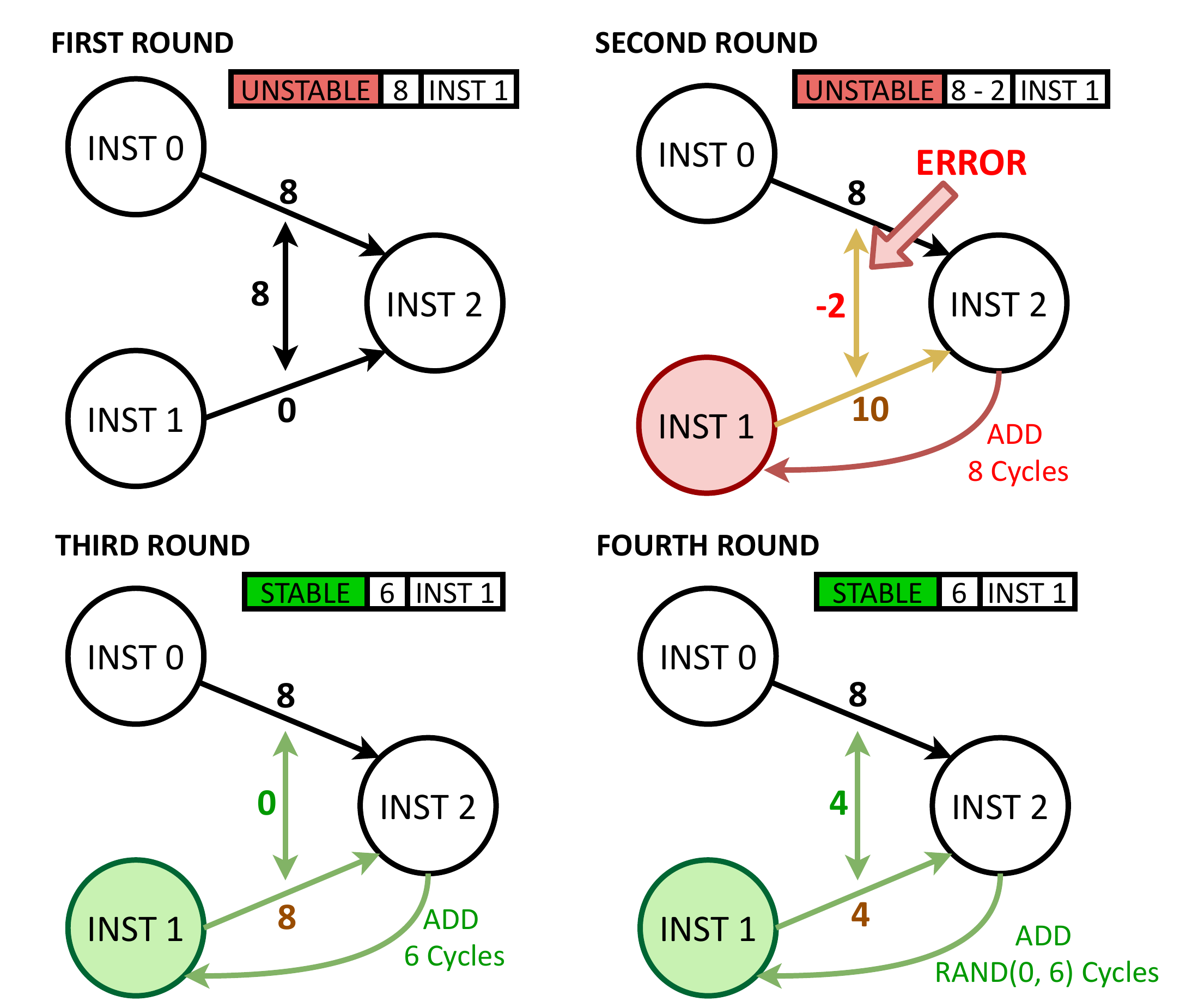}
    \caption{Slack detection and delay injection example.}
    \label{fig:ip}
\end{figure}

Figure~\ref{fig:ip} shows an example of how we detect the criticality and slack of instructions and how we inject a delay on non-critical instructions to desynchronize the execution. When INST 2 is issued in the first round, we detect the completion time for each of its input operands and using Equation~\ref{eq:slk} we calculate the slack (8 cycles). INST 0 is the critical instruction that we can't delay as it finished last and INST 1 is marked as non-critical with a slack of 8 cycles. Because INST 1 is still unstable, in its next appearance (second round) we inject it with the slack time as is. We expect that the new slack of INST 2 should now be 0 as INST 1 still needs 8 cycles to complete. However, practically we can find that INST 1 took 10 cycles to finish in this round (instead of the 8 cycles that we injected as a delay). The reason this happens is the backward dependencies of INST 1 to previous instructions that might have also been delayed. This is a common occurrence for all instructions that appear for the first time. This unstable delay slack injection occurs during the \textit{unstable phase}. %

During this unstable phase, %
the instruction is detected as unstable and the Slack Unit records the slack of INST 1 using Formula~\ref{eq:upd}. 
\begin{equation}
   slack_{update}^i =  slack_{old}^i - slack_{new}^i 
\label{eq:upd}
\end{equation}
The new slack of instruction $i$ ($slack_{update}^i$) will be the difference between the old slack ($slack_{old}^i$) and the new ($slack_{new}^i$). When the unstable phase is over in the third round, i.e., the injected delay will not change the criticality of the consumer (INST 2 in Figure~\ref{fig:ip}) anymore, the instruction will be marked as stable and the injected delay thereafter (fourth round) will be randomized with the updated slack being the upper bound of the random delay. This randomization introduces more noise into the power measurements, thus increasing the security of the system.

\section[Security Evaluation Methodology]{Security Evaluation\\Methodology}
\label{sec:security}

In this section, we provide a detailed description of the security evaluation used in this work.
First, we briefly introduce the power analysis attack for the AES algorithm.
Next, we discuss the dangers of using simple pass-fail security evaluation methods that many previous studies base their work upon. %
Finally, we describe the security framework and metrics used for analysis of our proposed design. %

\subsection{Power Analysis Attacks}

One of the goals of a power analysis attack is to reveal the secret keys used in encryption algorithms by observing the power consumption of a processor. To accomplish this, the adversary collects a large set of power traces from the core during encryption processing to examine and detect data and power dependencies.
In this section, we take a look at the AES-128 algorithm as an example of how to recover secret keys from a power trace.

In the first round of the AES-128 algorithm, a plaintext (in this example, a 16-byte input to the algorithm that will be encrypted) is loaded byte by byte and XOR-ed with the secret key to form the initial state of the ciphertext (the 16-byte encrypted output of the algorithm). Equation~\ref{eq:sbox} shows the SubByte step that is applied to the initial state of the ciphertext, which is also called an Sbox operation.
\begin{equation}
    I_{n} = Sbox[X_{n} \oplus K_{n}]
\label{eq:sbox}
\end{equation}
In Equation~\ref{eq:sbox}, $I_{n}$ is the $n^{th}$ byte of the intermediate value after the Sbox operation, $X_{n}$ is the $n^{th}$ byte of the plaintext, and $K_{n}$ is the $n^{th}$ byte of the secret key. The Sbox is a look-up table that takes a byte as input and substitutes it with another byte. For this attack, the Sbox is public and the plaintext ($X_{n}$) is known to the adversary.

\begin{figure}[t]
    \centering
    \includegraphics[width=0.95\linewidth]{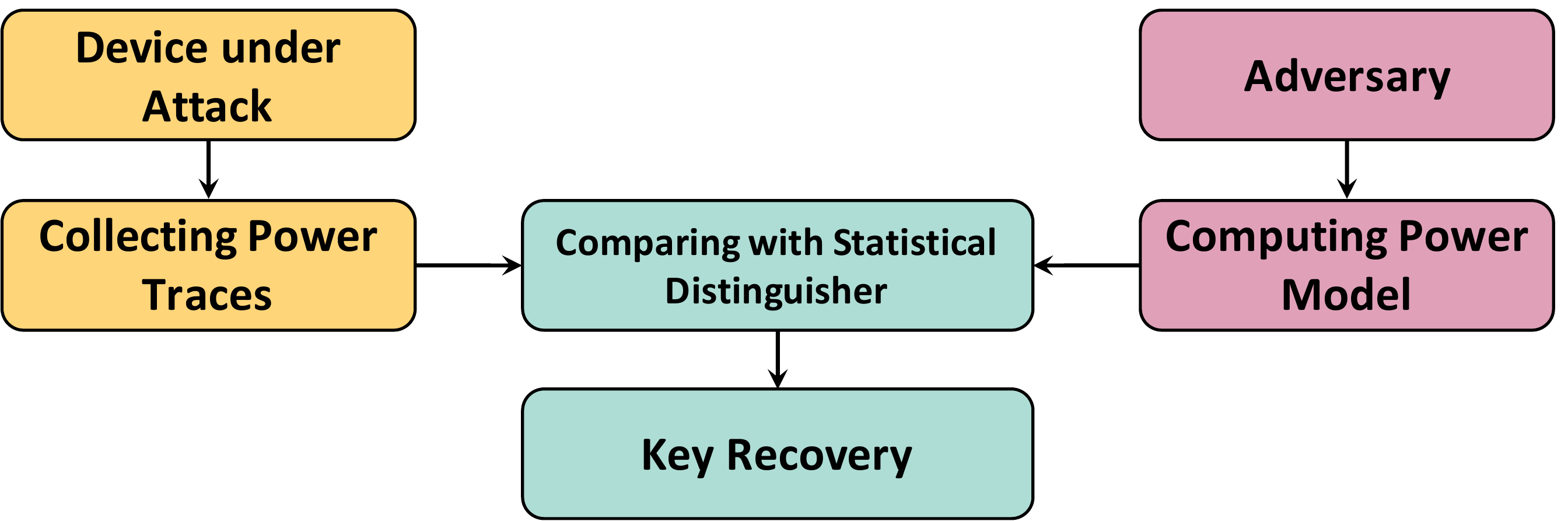}
    \caption{Flow of Differential Power Analysis (DPA) attacks. Power traces are collected from the device under attack and the adversary tries to compute the power model. A statistical distinguisher (like correlation, difference of means, etc.) is used to recover the secret key.}
    \label{fig:dpa}
\end{figure}

The overall modus operandi of a typical power analysis attack targeting a small part of the key or subkey\footnote{Typically, for AES, the subkey is one byte, where each byte of the key is attacked independently.} is shown in Figure~\ref{fig:dpa}. First, the adversary needs to get a power model reflecting the power consumption behavior of the device. This can be done using a priori assumptions on leakage behavior like Hamming weight/distance model~\cite{brier2004correlation} (better known as unprofiled attack), or by trying to characterize the actual model (i.e. profiled attack), using, for example, Gaussian template or supervised machine learning. Using this model, the adversary only needs to try all $2^8$ possible values for the secret key byte and examine if the computed intermediate value has a high statistical dependency with the collected power measurements. The attack outputs a probability for each of the $2^8$ guesses to be correct, and is successful if the most likely one corresponds to the actual key. This effectively reduces the brute force attack complexity from $2^{128}$ to $ 2^{12} (16~bytes \times 2^8)$ when using a divide and conquer methodology.
Many different statistical distinguishers, or attack tools, have been introduced in literature, with Kocher's Differential Power Analysis (DPA~\cite{kocher1999differential}), Correlation Power Analysis (CPA~\cite{brier2004correlation}), Mutual Information Analysis (MIA~\cite{batina2011mutual}) and the maximum likelihood template~\cite{chari2002template,schindler2005stochastic} being some examples. %
In the rest of Section~\ref{sec:security}, and independently of which statistical distinguisher is used, we will refer to DPA as any attack taking advantage of varying plaintext. This includes, for example, the aforementioned DPA and CPA.

\subsection{Limitations of Simple Pass-Fail Tests}

\noindent \textbf{T-test:} 
The order of leakage is defined by the statistical moment in which the meaningful information depends. For example, first-order (respectively second-order) leakages extract information in the mean (respectively variance/co-variance). %
A first-order secure masking countermeasure ensures that no information lies in the mean of side-channel traces, forcing an adversary to extract information in (at least) the variance and co-variance.      
The Test Vector Leakage Assessment (TVLA~\cite{gilbert2011testing}) methodology was introduced to detect the presence of this first-order leakage for masked implementations, and can be extended for higher orders~\cite{tvla}. %
That is, a T-test failure shows the existence of such leakage, while a success can only state that no such leakage was found for a given number of measurements. %
However, it has since often been used as a pass-fail method to evaluate the resilience of an implementation, where the hidden assumption is that a T-test failure with a total of $Q$ measurements roughly translates to security up to $Q$ measurements. However, such an assumption has been shown to be incorrect~\cite{standaert2018not}. That is, the standard T-test should only be used to show the presence of first-order leakage when the null hypothesis is rejected. However, no conclusion should be made when it is not the case. As a direct consequence, using the T-test for the \textit{hiding} style of countermeasure, where leakages would exist at first-order, should be avoided. %

\medskip

\noindent \textbf{Fail/success unprofiled CPA:} The strengths and weaknesses of unprofiled CPA lies in its assumptions. 
On one hand, it is designed to process one time sample at a time (also known as univariate leakage~\cite{standaert2018not}). While one can combine time samples together before performing CPA (i.e. multivariate attack~\cite{standaert2018not}), doing so is sub-optimal. 
On the other hand, CPA assumes some leakage model. For these reasons, CPA is often a good security estimate for unprotected implementations where the side-channel traces are well aligned and where the power model is well known (e.g. Hamming weight/distance). However, this no longer holds when the type of leakage diverge from these assumptions. That is, an implementation that changes the underlying leakage model (such as power balancing~\cite{hwang2006aes}), or that introduces jitter or misalignement from one measurement to another, will drastically drift apart from these hypotheses and subsequently makes CPA sub-optimal. In other words, \textit{a failed CPA under sub-optimal assumptions does not provide a sound security evaluation}.  A direct result of the use of this method will be a false sense of security (a higher-than expected security result), while a more appropriate attack (which exploits a well characterized leakage model or apply measurement alignment/processing techniques) might achieve key recovery with significantly fewer observations. %
In the next subsection, we propose an evaluation framework which also considers powerful adversary capable of applying the right leakage model and processing techniques to exploit the available leakage in an optimal manner.

\subsection{Proposed  Evaluation Framework}

The limitations of the previously discussed methods can be summarized as using weak or incorrect adversarial assumptions. As previously shown in Figure~\ref{fig:dpa}, the two main ingredients for a power attack are the leakage model and the statistical distinguisher used. The further the model is from the actual leakages from the device, the worse the attack will be. This issue typically happens when using a Hamming model for power balancing, for example, which inherently changes the leakage model. On the other hand, as methods such as CPA are not meant to combine leakages together, they are less optimal for multivariate attacks compared to template attacks~\cite{chari2002template}. As a result, evaluating the security of a device with the wrong model and method can lead to a false sense of security, while another method could succeed using fewer measurements. This can be mitigated by considering stronger adversarial capabilities, eventually up to a potentially non-existent (extremely strong) adversary in order to approach a lower-bound (conservative) estimate of security guarantees.

In this work, we will illustrate this by providing different levels of security analysis for three types of adversaries, that we call \textit{\basic}, \textit{\educated}, and \textit{\advanced}. First, in order to compare our work to existing literature, the \basic{} evaluation considers an adversary that simply applies standard CPA with a Hamming weight model on the Sbox output~\cite{brier2004correlation}. Second, as a part of the security brought by our countermeasure comes from desynchronization, an \educated{} evaluation will be performed with a CPA analysis with the same model, additionally pre-processing traces to defeat countermeasures with alignment techniques such as integrating the leakage over different time samples. This simple method aims to show how basic knowledge of the implementation and a simple attack twist can greatly change the evaluation outcome (effectively demonstrating a lower security guarantee). Finally, the \advanced{} evaluation aims to approach the lower security bound (a conservative security estimate) by assuming an adversary adopting profiled attacks. That is, we will first take advantage of a profiling (or training) set in order to mount a multivariate template attack~\cite{chari2002template} augmented with profiled Principal Component Analysis (PCA) for dimensionality reduction~\cite{archambeau2006template}.
PCA applies a linear transformation that projects high-dimensional
data into a low-dimensional space while preserving the data variance, by computing the eigenvectors of the co-variance matrix.
The training or profiling phase with PCA allows an adversary to learn the precise leakage model and better characterize the underlying countermeasure, leading to a confident lower-bound on security.

\subsection{Security Metrics}

As the divide-and-conquer approach allows one to attack each byte of the key independently, %
we target an attack with only one key byte without loss of generality.
For all three attack methodologies, the resulting vector of 256 probabilities/scores for each key guess is denoted by $\mathbf{p}$. In our evaluation, we will compute the security using the following metrics:

\medskip

\noindent \textbf{Key rank (byte):} Given the probability/score  vector $\mathbf{p}$ resulting from an attack, the rank of the key (byte) is given by the number of key candidates having a higher probability than the correct key.

\medskip

\noindent \textbf{Guessing entropy~\cite{standaert2009unified}:} For a given key byte $k$, the guessing entropy (GE) is the average key byte rank within its vector of probability $\mathbf{p}$. We define by $\mathsf{rank}(\mathbf{p},k)$ the function that returns the rank of the subkey $k$ within the vector $\mathbf{p}$.  From a set of $n_a$ independent attack result vectors $\mathbf{p}_i$, Equation~\ref{eq:ge} allows one to compute the guessing entropy.

\begin{equation}\label{eq:ge}
\text{GE} = \frac{\sum_{i=0}^{n_a-1} \mathsf{rank}(\mathbf{p}_i,k)}{n_a} \cdot
\end{equation}

\medskip

The use of guessing entropy provides more information than the commonly used measurement to disclosure (MtD) metric \cite{mangard2004hardware}. 
First, it provides averaged information over several independent attacks. This minimizes the over- and under-estimation of the actual security, as a single experiment could be an outlying result. Second, it additionally shows the global key recovery progression instead of only reporting the overall number of traces. Indeed, as these attacks belong to the class of divide-and-conquer methodologies, one can trade-off side-channel complexity for a computational one and recover the key through enumeration before the rank reaches one~\cite{poussier2016simple}. As a result, only looking at the number of measurements required to recover the key can be misleading, as the implementation might be broken with fewer traces with some brute-force or key enumeration. %

\section{Experimental Setup}
We implement \name{} on top of SonicBOOM~\cite{zhaosonicboom}, an open-source RISC-V out-of-order processor. 
The configuration of the processor is shown in Table~\ref{tb:config}. We leverage Galois Linear Feedback Shift Register (GaloisLFSR) function~\cite{press2007numerical, marsaglia2003xorshift} provided by Chisel~\cite{bachrach2012chisel} to randomize the stable delay. The system time when we generated the core and the runtime clock cycles are used together to generate a random seed. %
This random seed is more secure than the default seed (only runtime clock cycles) as it is almost impossible for the adversary to know when the core was generated.

\begin{table}[tb]
\caption{System configuration.}
\centering
\resizebox{0.99\linewidth}{!}{%
\begin{tabular}{@{}l|l||l|l@{}}
\toprule
\textbf{Parameter}                & \textbf{Value} & \textbf{Parameter}                & \textbf{Value}     \\
\midrule
RF/Fetch Buffer/ROB             & 128/24/96 entries   & L2 Cache            & 4 MB, 8-way set assoc.     \\
Issue Queue              & 3$\times$8 entries & Bus Protocol             & AXI  \\
Execution Units          & 5 (1 MEM, 3 ALUs, 1 FPU)   & *(\name{}) DT & 208 B, 4-way set assoc.     \\
Branch Predictor        & Next-line, backing predictor  & *(\name{}) CT & 144 B, 4-way set assoc.      \\
Cache line size          & 64 B   & *(\name{}) NCT & 192 B, 4-way set assoc.    \\
L1-I and L1-D Caches          & 32 KB, 8-way set assoc.  & &   \\
\bottomrule
\end{tabular}}
\label{tb:config}
\end{table}

Apart from the baseline (SonicBoom) and \name{}, in this paper, we also implement three generic processors to compare with the \name{} implementation. For each of these processors, we naively inject random delays (up to 8 cycles) by a given probability to match either the performance of our \name{} implementation (in the random-iso-performance configuration), or the level of security protection of our implementation (in the random-iso-security design). The detailed configuration parameters of these processors and all other processors that are used in security and performance evaluation are described in Table~\ref{tb:config} and Table~\ref{tb:dsp}.

\textbf{Benchmarks}. To determine the exact performance degradation for the encryption application used in this study, we run the AES-128 encryption engine with 2,000 plaintexts both on SonicBoom and \name{}. We then run the micro-benchmarks provided by Chipyard~\cite{chipyard} on the out-of-order cores from Table~\ref{tb:dsp} to evaluate the overheads as introduced by our protection scheme and the random-delay injection schemes
for a set of general-purpose applications. %
The micro-benchmarks consist of a basic set of applications designed to test the functionality of the processor in different scenarios, e.g., complex matrix computation, multi-threaded applications, sorting, Dhrystone, etc. Finally, we boot a full Linux system on each core using FireSim~\cite{karandikar-firesim-isca18} to run the the SPEC CPU2017 benchmark suite. This allows us to observe the performance of different processors in a more general environment. However, because of limitations in FireSim, we support only 10 SPEC CPU2017 benchmarks.

\textbf{Area and Power}. We leverage the Synopsys Design Compiler (DC) to synthesize the \name{} and the SonicBoom, both of which are generated by the default synthesizable SonicBoom configuration. Next, we use VCS to perform the gate-level simulation on synthesized processors to generate realistic gate activity. PrimePower is then used to generate the power consumption of the processor by analyzing the gate-level waveform and gate-level code.

\begin{table}[t]
\caption{Implementation of different processors.}
\resizebox{1.0\linewidth}{!}{%
\begin{tabular}{l|c|c}
\toprule
\multicolumn{1}{c|}{\textbf{Processor Name}}  & \textbf{Platform}                 & \textbf{Description}                                                                              \\ 
\midrule
ooo-baseline     & SonicBoom & Unprotected baseline out-of-order processor                                                            \\ \hline
io-baseline     & Rocket chip &  Unprotected in-order processor                                                            \\ \hline
\name{}     & \makecell[c]{SonicBoom +\\ Slack Unit}  & Secure instruction scheduling processor                                                           \\ \hline
random-iso-perf       & \makecell[c]{SonicBoom +\\ random delay}    & \makecell[c]{Delay injection probability is 5\%\\ Try to match the performance of \name{}} \\ \hline
random-iso-security   &    \makecell[c]{SonicBoom +\\ random delay}   & \makecell[c]{Delay injection probability is 20\%\\ Try to match the \advanced{} security evaluation of \name{}}    \\ \hline
random-aggressive &  \makecell[c]{SonicBoom +\\ random delay} & \makecell[c]{Inject random delay for all instructions\\ Naive and aggressive implementation}         \\ 
\bottomrule
\end{tabular}}
\label{tb:dsp}
\end{table}

\textbf{Power traces simulation framework}. 
We use a Hamming weight leakage model to generate power traces. Equation~\ref{eq:model} shows the estimated power at any given time of the execution.
\begin{equation}
   \mathsf{power}(T) = 
   \begin{cases}
       0       & \nexists \quad \mathsf{inst}: \mathsf{inst}.\mathsf{WB} = T\\
    \displaystyle\sum_{\mathsf{inst}}\mathsf{HW}(\mathsf{inst})  & \forall \quad \mathsf{inst}: \mathsf{inst}.\mathsf{WB} = T
   \end{cases}
\label{eq:model}
\end{equation}
The write-back time of the instruction $\mathsf{inst}$ to the register file is denoted by $\mathsf{inst}.\mathsf{WB}$. Also, $\mathsf{HW}(\mathsf{inst})$ is the Hamming weight of the data that instruction $\mathsf{inst}$ writes to the register file. We use the behavioral simulation of the SonicBoom core to gather this information and generate the power traces. We only use the updates to the register file to generate the power traces since the register file in SonicBoom consumes much more power than system and memory bus (33$\times$ more power consumption for integer register file compared to system and memory bus, and 61$\times$ more power consumption for integer+floating-point RF).

To perform security evaluations, each implementation consists of two sets of one million traces each. The first set (attack set), is composed of a fixed key and randomly varying plaintext. The second set (profiling set), is composed of randomly varying keys and plaintext that are known by the adversary to perform \advanced{} profiling methods.

\section{Experimental Results}
\label{sec:evaluation}

\subsection{Security Evaluation}
\label{sec:security_evaluation}
To compare our work to the literature, we perform a \basic{} evaluation, which performs a regular CPA without any modifications. Next, we assume an adversary that knows the desynchronization aspect of our countermeasure in order to perform the \educated{} evaluation. The adversary will now use integration over the time samples to reduce the signal reduction coming from the desynchronization. This will show that basic insight and twist from the adversary can greatly change the security evaluation outcome and give a false sense of security. Finally, we present results of our \advanced{} evaluation that aims to approach the lower security bound by assuming a strong profiled adversary. For each method we compare the security benefits of all cores with their corresponding ooo-baseline with respect to the same attack.

\subsubsection{\Basic{} evaluation}

\begin{figure}[tb]%
\begin{center}
\includegraphics[width=0.9\linewidth]{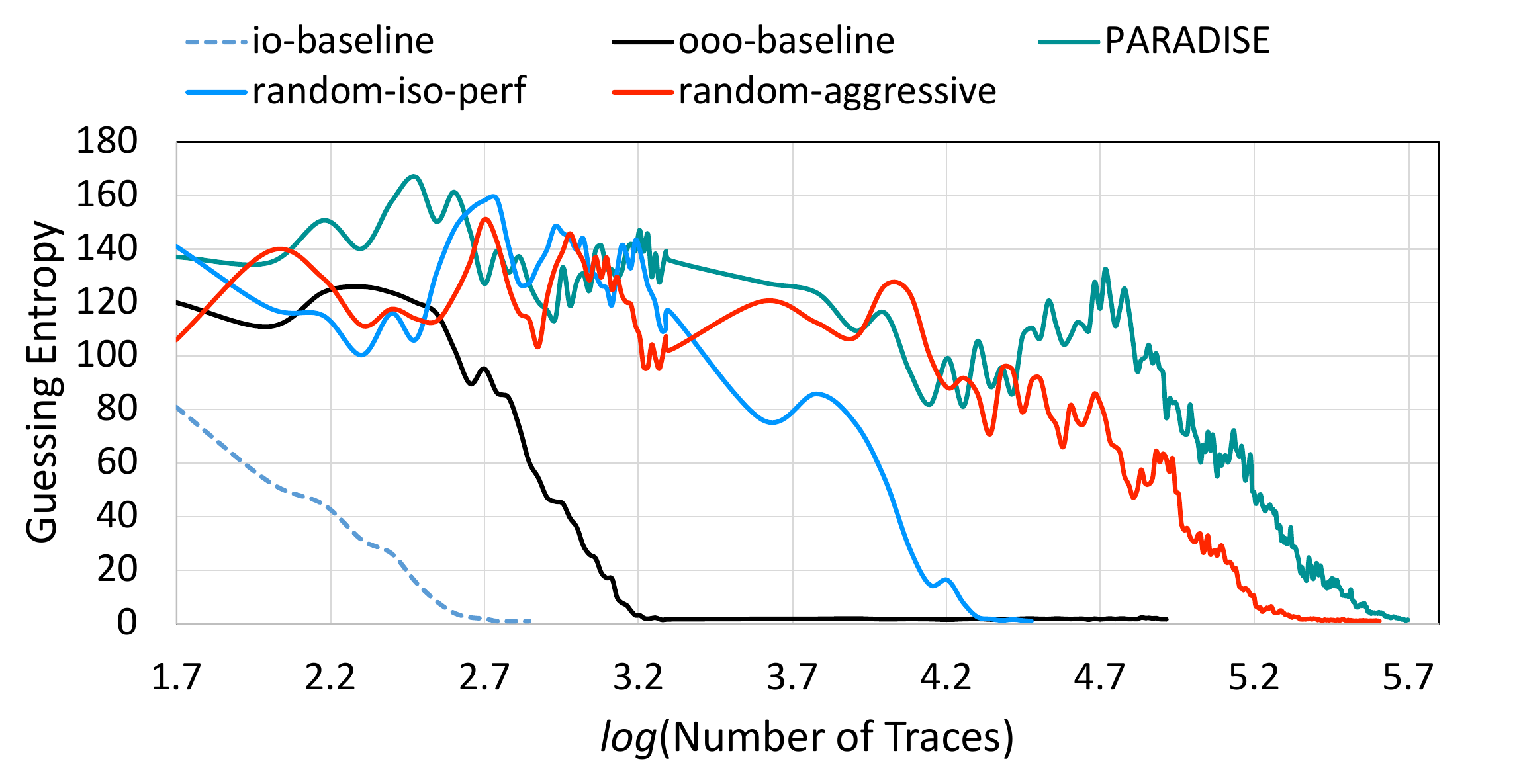}
\caption{Results of the \basic{} security evaluation.  The x-axis corresponds to the number of traces, and the y-axis corresponds to guessing entropy. 
}
\label{fig:naive}
\end{center}
\end{figure}

As a first analysis, we divided our 10 million attack traces into 20 subsets, and performed a standard CPA on each of them and average them to compute the guessing entropy. The results are shown in Figure~\ref{fig:naive}, and will be the main point of comparison with other works. The x-axis corresponds to the number of traces and the y-axis corresponds to guessing entropy. A guessing entropy of $0$ indicates that the correct key is highest ranked (on average) and thus the attack is successful.

As a first observation, we can observe the gap between \rocket{} and \baseline{} sets, which respectively recovers the key with 500 and 1,800 traces. This shows the simple security benefit of using an out-of-order core instead of in-order ones. Indeed, out-of-order core still provides some randomness in the computation timings with a small security benefit. As \baseline{} corresponds to the unprotected implementation on our out-of-order case study, we will use it as a reference for our security benefits. First, we can see that the \secure{} version of our countermeasure increases the number of traces needed to 470,000 when using standard CPA. When considering \basic{} evaluation, this shows a security of \parab{} \baseline{}.
However, using random delays with the same performance as \secure{}, shown by the \isoperf{} performances, only requires 22,000 traces for key recovery, only corresponding to a benefit of \isoperb{}. This shows that, when considering standard CPA, our method shows greater security benefits than random delays with overheads. Finally, \aggressive{} only requires 220,000 traces, which corresponds to a security benefit of \aggb{}, and is less than the \secure{}  implementation. While this can look surprising at first, it can be explained when looking at the two security benefits brought by our countermeasure. As explained in Section~\ref{sec:motivation}, the security mainly comes from (1) desynchronization and (2) more randomness in the register's content. The \aggressive{} implementation focuses on increasing the desynchronization, without much change in the register content randomization, which is higher in the case of the \secure{} one. %

\subsubsection{\Educated{} evaluation}

\begin{figure}%
\begin{center}
\includegraphics[width=0.9\linewidth]{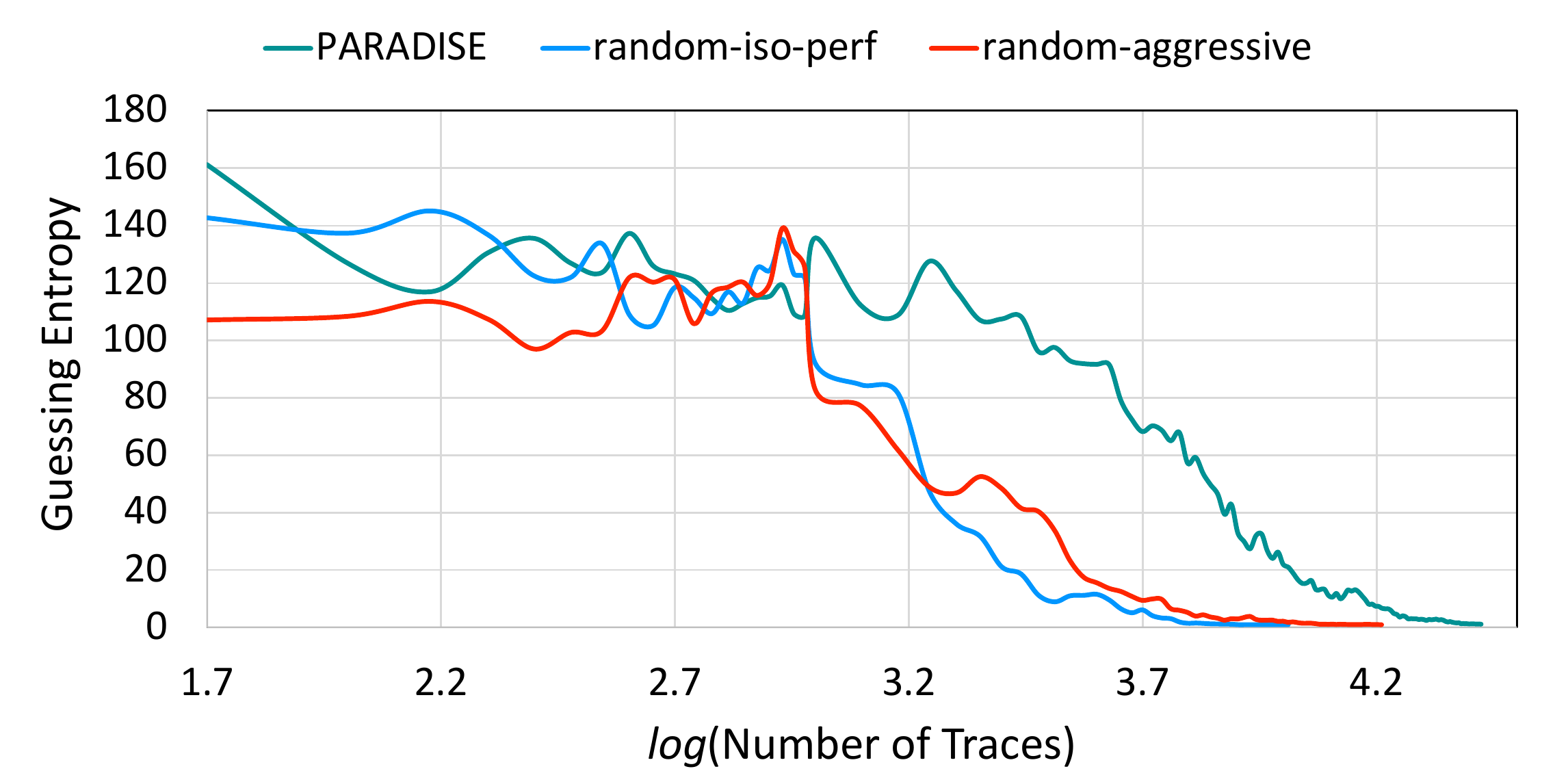}
\caption{Results of the \educated{} security evaluation. The x-axis corresponds to the number of traces, and the y-axis corresponds to guessing entropy. 
}
\label{fig:educated}
\end{center}
\end{figure}

As a second evaluation, we illustrate how some basic knowledge on the type of implemented countermeasure, and a simple optimization of the attack itself can drastically change the outcome. 
As a big part of security brought by our countermeasure is brought by desynchronization, standard CPA being a univariate attack is inherently suboptimal. Instead, we now assume an adversary with some insight that simply combines sets of $N$ consecutive time samples on the trace together (time integration) prior to performing the CPA to reduce the effect of desynchronization. For each implementation, we tested values of $N={20,50,100,150,200}$, for which the best corresponding results are shown in Figure~\ref{fig:educated}. Again, the x-axis corresponds to the number of traces and the y-axis represents guessing entropy.

We do not report any results for \rocket{} and \baseline{} implementations, as \basic{} evaluation was better (in term of number of traces required) for non-existent and limited desyncronization in \rocket{} and \baseline{} respectively.  %
For that reason, we will still use the numbers from the \basic{} evaluation for these two implementations. However, we can see that the \educated{} evaluation drastically reduces the required nu ()mber of traces for both \secure{} and \aggressive{} implementations, which now respectively broken with 22,250 ($12\times$) and 20,000 ($11\times$) traces instead of 470,000 ($261\times$) and 220,000 ($122\times$) respectively. This simple attack optimization shows the danger of using sub-optimal attack strategies for security evaluations. %
Interestingly, minor benefits are seen for \isoperf{} implementation where 21,000 traces are now required instead of 22,000. %

\subsubsection{\Advanced{} evaluation}

\begin{figure}[tb]%
\begin{center}
\includegraphics[width=\linewidth]{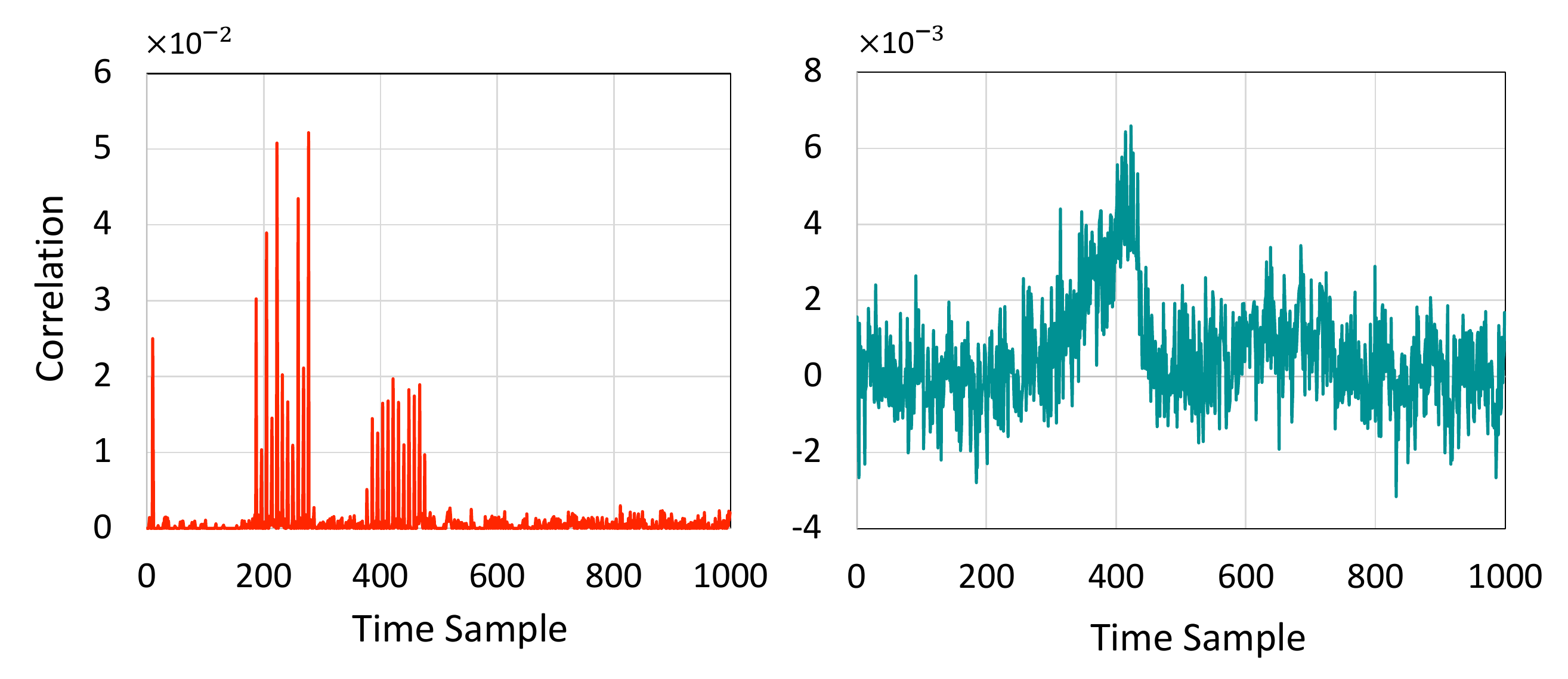}
\caption{Leaking regions using profiled CPA for \baseline{} (left) and \aggressive{} (right). The x-axis corresponds to the time samples, and the y-axis corresponds to correlation coefficient.}
\label{fig:rkey_corr}
\end{center}
\end{figure}

Our last evaluation considers a powerful adversary being able to profile the leakages using, for example, a copy or clone of the device under attack for which she has complete control. First, we use profiled CPA~\cite{durvaux2016improved} in order to identify leaking features in the trace. The results are shown in Figure~\ref{fig:rkey_corr}, where the left part shows the results for the \baseline{} and the right part shows results for \aggressive{} implementations. 
The x-axis corresponds to the time samples, while the y-axis shows the correlation coefficient.

As we can see, leakages are clearly identified with several peaks for the \baseline{} implementation. We observed similar behavior for the \rocket{} and \isoperf{} one. For these two implementations, we thus selected all time samples having a correlation above $0.005$ as valid attack samples. However, the leaking samples for the \aggressive{} implementation are less clearly identified, as shown by the Gaussian shape correlation trace covering around 200 time samples. This was similarly observed for \secure{} and \isoperf{} implementations, which is the result of the desynchronization brought by the countermeasures. In that case, we selected all time samples happening before and after the peak regions as valid.

\begin{figure}%
\begin{center}
\includegraphics[width=0.9\linewidth]{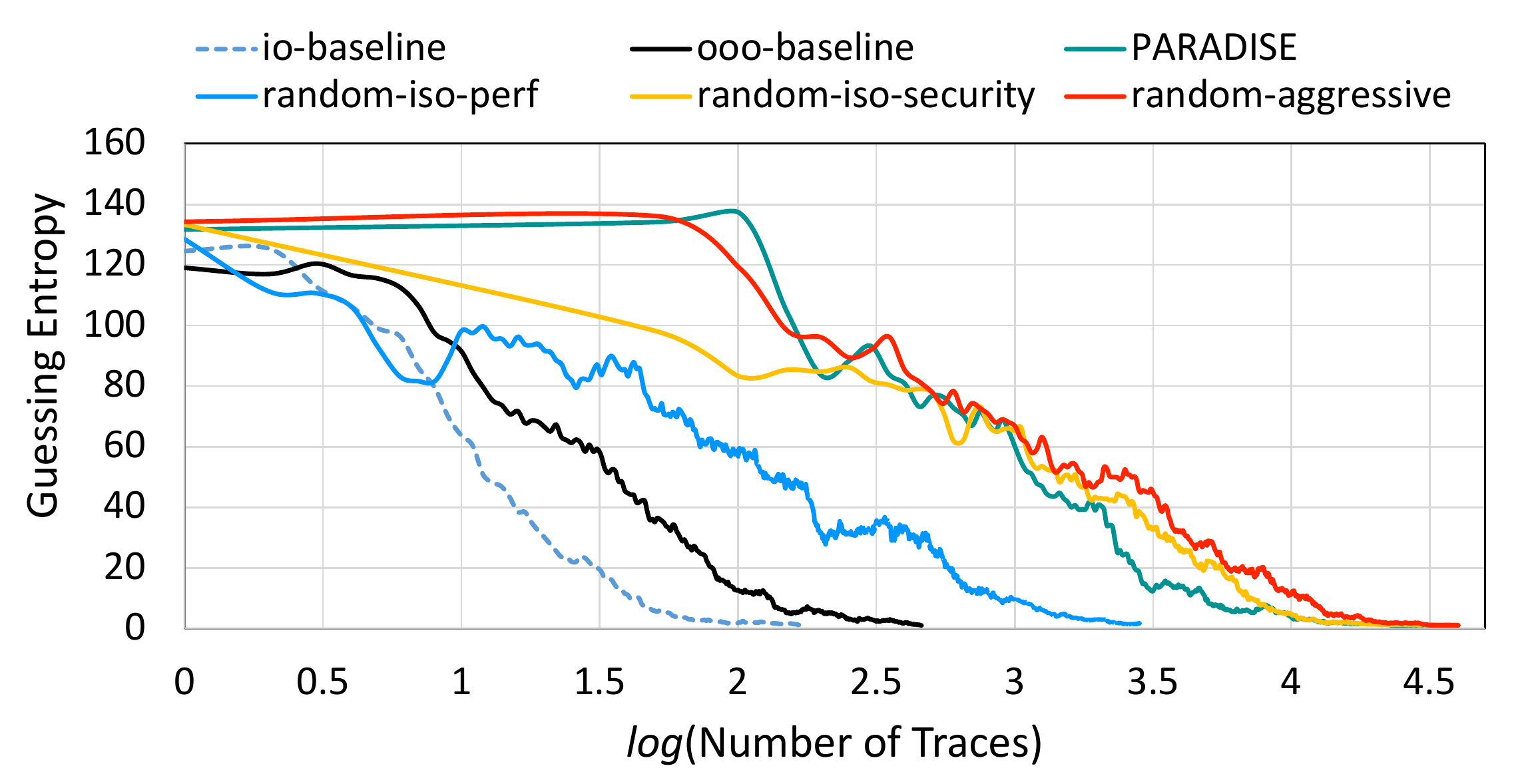}
\caption{Results of the \advanced{} security evaluation.  The x-axis corresponds to the number of traces, and the y-axis corresponds to guessing entropy.}
\label{fig:advanced}
\end{center}
\end{figure}

Once the points of interest are selected, we perform a profiled Principal Component Analysis (PCA)~\cite{archambeau2006template} in order to further reduce the dimensionality and to project the sample into a more informative space. The result of the projection is then fed into a multivariate template attack~\cite{chari2002template}. For each implementation, we use a different number of principal components, and show the best results for each of them in Figure~\ref{fig:advanced}. We additionally show the results for the \isosecu{} implementation, having similar security performances as the \secure{} one (tailored specifically in the case of the \advanced{} evaluation).

We can see that the \advanced{} method produces better results in term of attack power than the \basic{} and \educated{} ones. 
First, \rocket{} implementation now only requires 125 traces instead of 500 when using standard CPA.
Second, \baseline{} implementation is now broken with only 400 traces, which we will now use as comparison reference as opposed to 1,800 for \basic{} evaluation. 
Next, \secure{} implementation now only requires 13,500 traces, showing a security gain of \paraa{}. However, \isoperf{} is now broken with 2,200 traces, hinting that our countermeasure is \isopera{} more secure than random delays when considering a strong adversary.
Interestingly, as opposed to previous results, the \aggressive{} implementation now requires 15,000 traces, which is more than the \secure{} one with a benefit of \agga{}. Indeed, as we now profile the leakage model, the effect of the vertical noise is now reduced, which has more impact on the \secure{} implementation.
Overall, this shows that using the wrong or too sub-optimal attack against a given countermeasure can lead to a false sense of security. Indeed, from standard CPA to multivariate templates, the number of required traces has been divided by 35 for the \secure{} implementations, and by 15 for the \aggressive{} one due to wrong model assumptions. However, the number of traces unprotected \baseline{} implementation was only divided by 4.5, as the model was already fitting more. From these observations, we highlight as a cautionary note when presenting results from two different implementations, that it can only reflect the adversary's assumptions. On that matter, it is more conservative to apply  powerful attacks, thus assuming a strong adversary.

\subsection{Performance Evaluation}
\label{sec:perf}

\begin{figure}[t]
    \centering
    \includegraphics[width=0.9\linewidth]{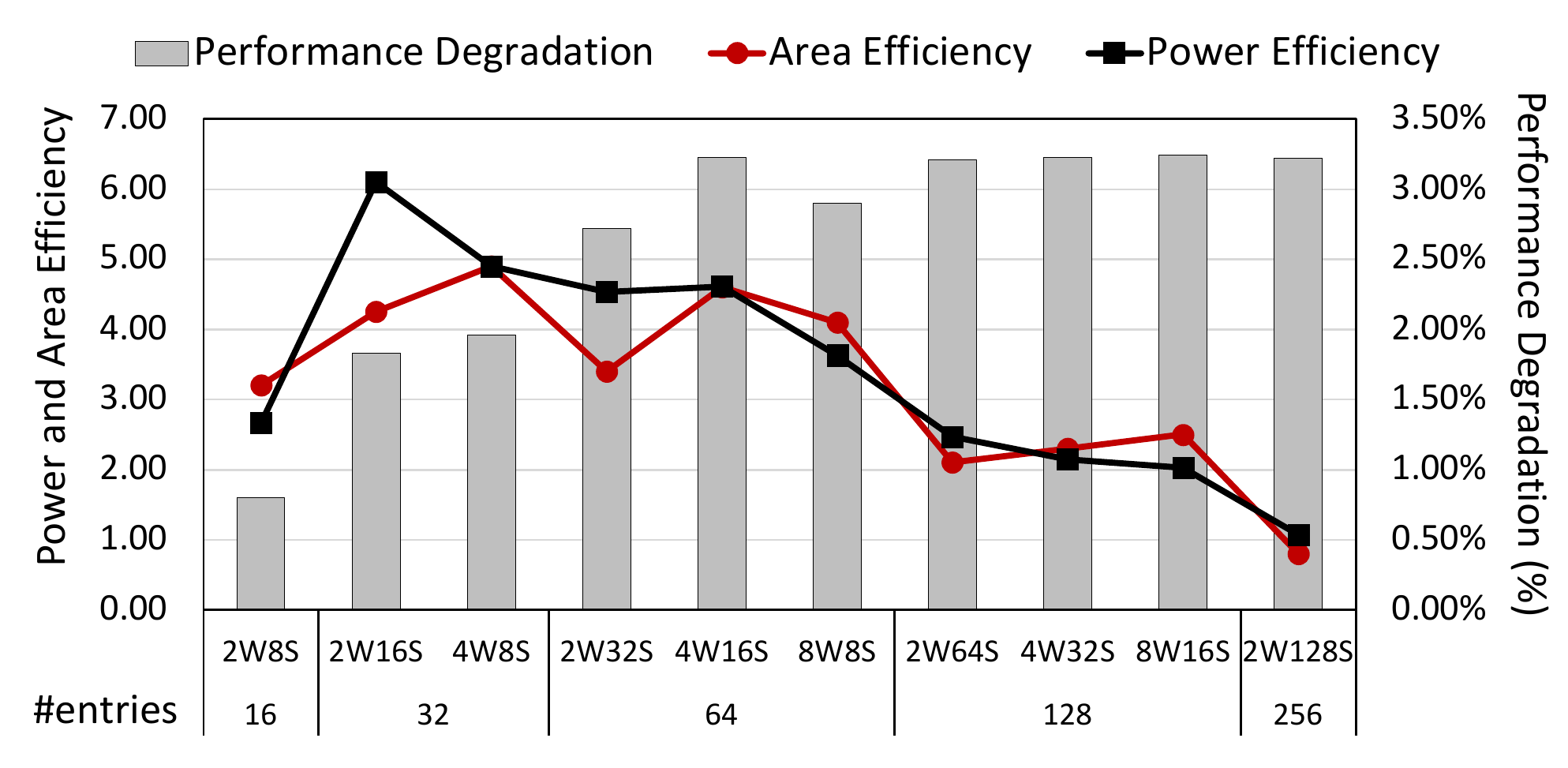}
    \caption{Performance degradation compared to ooo-baseline core and area/power efficiency using different parameters. $W$ and $S$ represent number of $ways$ and $sets$ respectively, the $\#entries$ are equal to $W\times S$, which denote the total number of $entries$ of each table in the Slack Unit.}
    \label{fig:aes_all}
\end{figure}

We evaluated several configurations of \name{} by running an AES-128 encryption engine with 2,000 plaintexts to find the best combination of performance, power and area. Figure~\ref{fig:aes_all} shows the best combination of power and area efficiency to be a 4-way and 16-set Slack Unit. Power-performance and area-performance efficiency shown in Figure~\ref{fig:aes_all} are calculated as $\frac{overhead_{perf}}{ overhead_{power}}$ and $\frac{overhead_{perf}}{ overhead_{area}}$. We use the highest number for $overhead_{perf}$ because in this scenario \name{} is going through a longer \textit{unstable phase} and although it injects more unstable delays, it collects more runtime information to improve in later iterations. Therefore, more instructions will be reordered and the desynchronization of the execution increases with the overall security improved. The rest of our performance results were taken with all structures in the Slack Unit being 4-way with 16 sets. 

\begin{figure}[t]
    \centering
    \begin{tabular}{@{}c@{}}
      \includegraphics[width=0.9\linewidth]{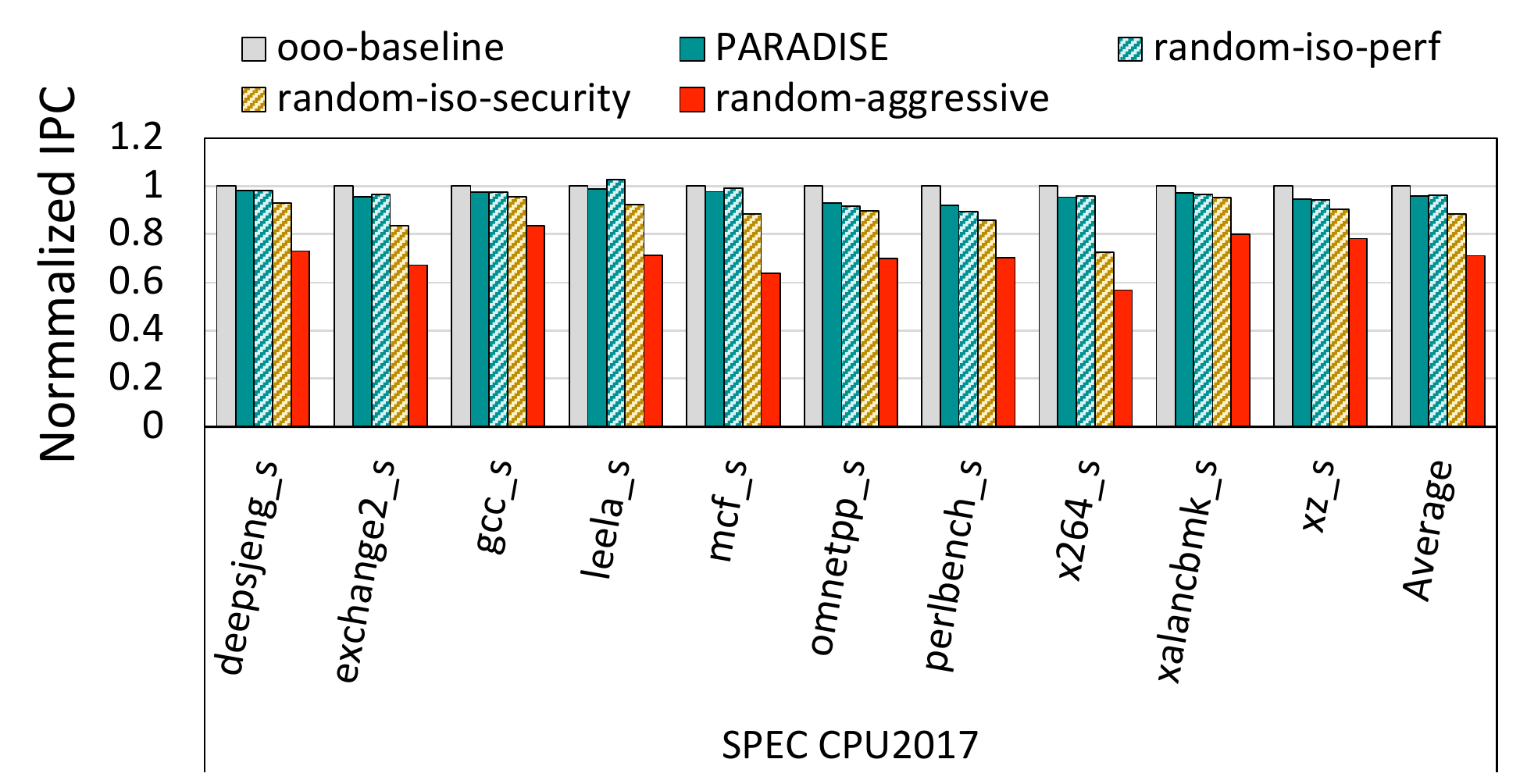}
    \end{tabular}
    
    \begin{tabular}{@{}c@{}}
      \includegraphics[width=0.9\linewidth]{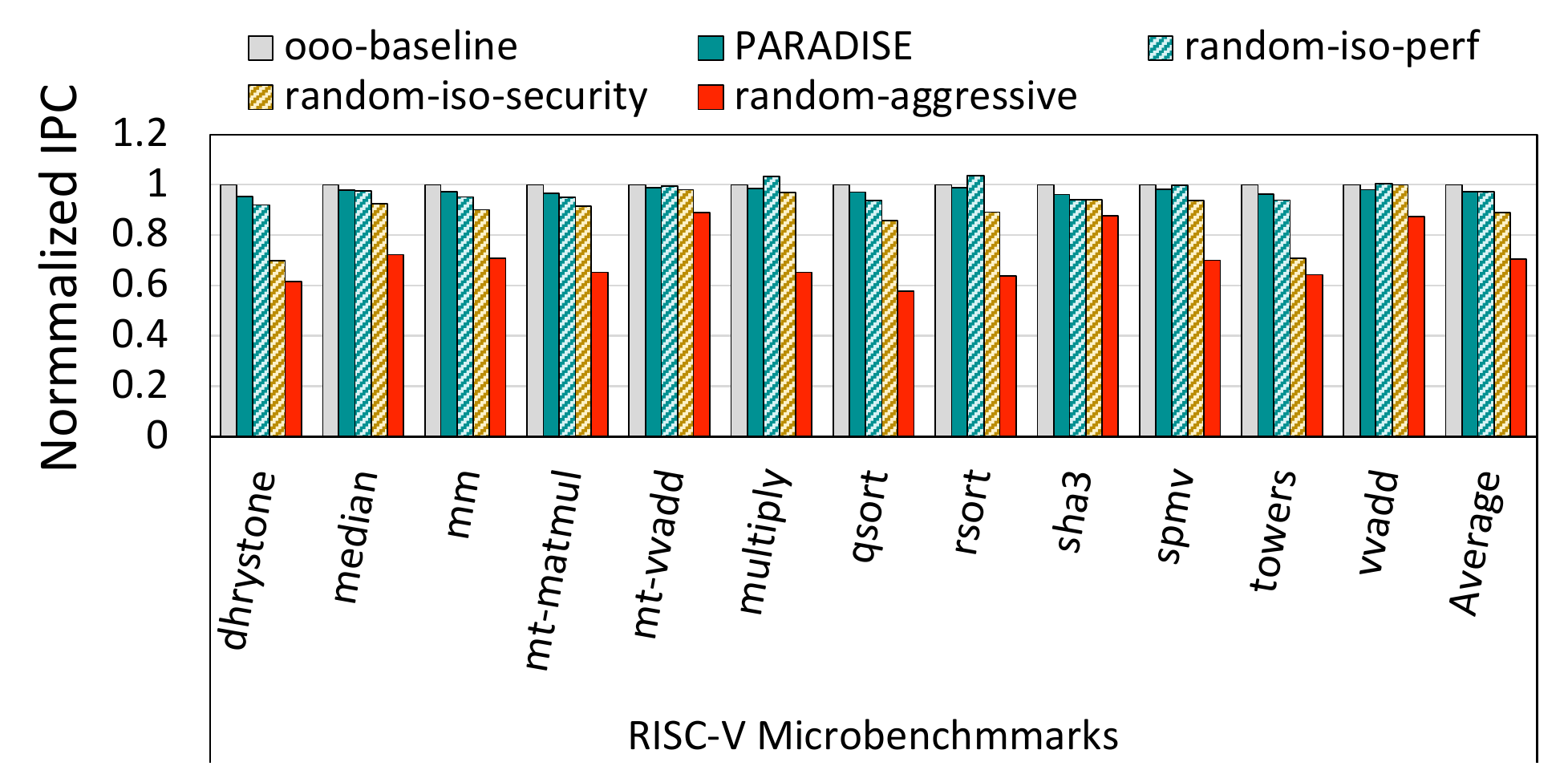}
    \end{tabular}
    \caption{Performance of different cores normalized to ooo-baseline core. *We run all SPEC CPU2017 benchmarks compatible with FireSim.}
    \label{fig:spec_perf}
\end{figure}

Figure~\ref{fig:spec_perf} shows the performance of \name{} (normalized IPC to the ooo-baseline) on two different sets of benchmarks. At the bottom, the RISC-V bare-metal microbenchmarks show an average overhead on \name{} of 2.6$\%$ (maximum is 4.8 $\%$). For the random-iso-perf, random-iso-security, and random-aggressive implementations we get 2.7\%, 11\%, and 28\% overheads respectively. However, these applications don't have a real software stack and some of them cannot generate a stable result due to the limited number of instructions. For this reason we evaluated a subset of SPEC CPU2017 benchmarks (top of Figure~\ref{fig:spec_perf}). We could only evaluate those applications that were compatible with FireSim. The average overhead of \name{} in this case is 4\%, while for random-iso-perf, random-iso-security, and random-aggressive it is 4.2\%, 12\%, and 29\% respectively. For all the applications evaluated, the average performance overhead of \name{} is 3.7\%. An in-depth analysis of the performance results showed that the overhead produced comes mostly from the \textit{unstable phase} were we inject an unstable delay. As soon as the criticality of the instructions and the slack become stable, after a few iterations of the algorithm, the performance of the processor returns to its baseline levels.

\begin{figure}[t]
    \centering
    \includegraphics[width=0.9\linewidth, trim= 0 1cm 1cm 0]{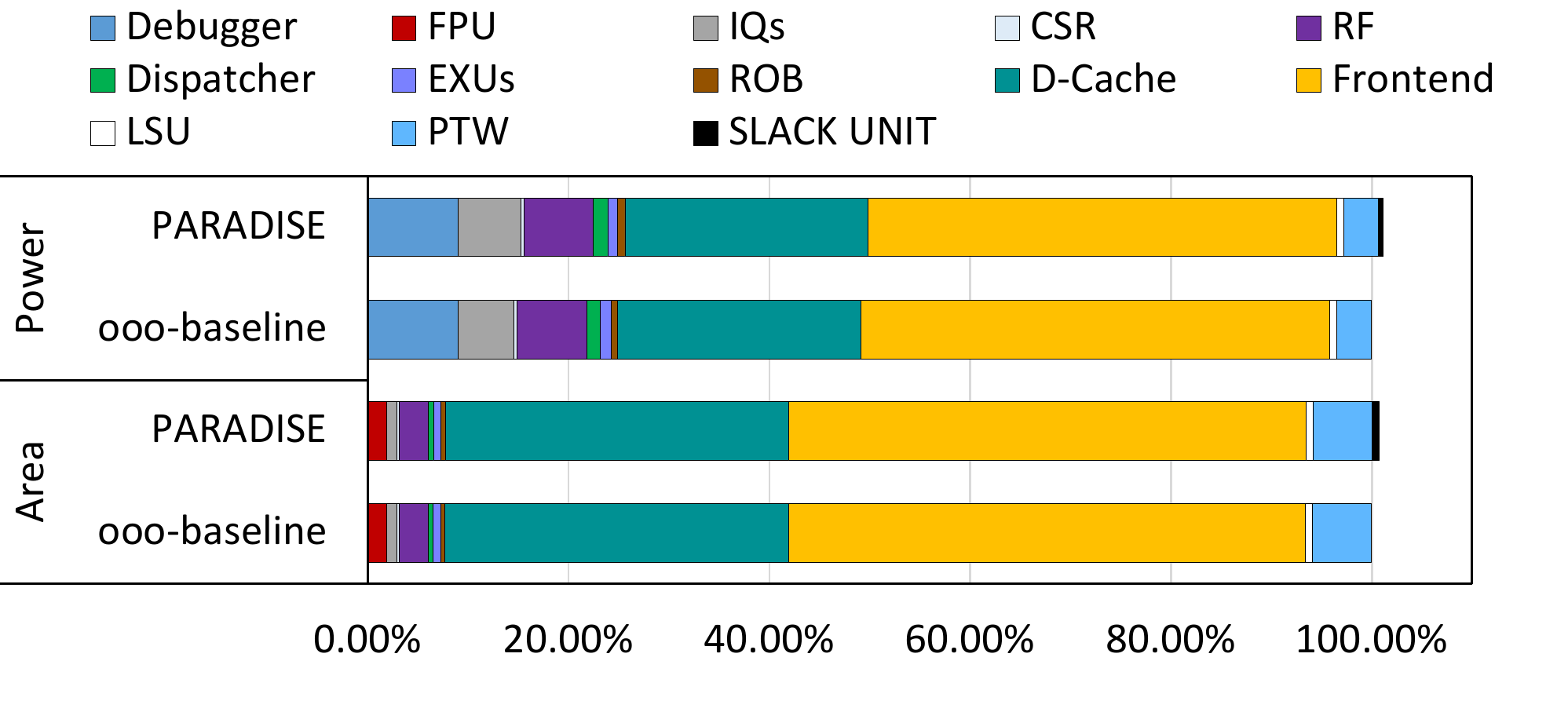}
    \caption{Power and area overheads of \name{}.}
    \label{fig:power_area}
\end{figure}

\begin{table*}[t]
\centering
\caption{Comparison of existing power attack countermeasures. *For each method we compare the security benefits of each core to their corresponding unprotected baseline with respect to the same attack. Values of other works are presented as reported; '-' for imprecise reporting or absence of reporting of the information. $\ddagger$ Evaluation reported as \textit{negative} and/or inconclusive tests that did not recover the key. 
}
\resizebox{1.0\linewidth}{!}{%
\begin{tabular}{l|c|c|c|c|c|c|c|c|c|c|c}
\toprule
\multirow{2}{*}{Paper}       
& Hardware & Algorithm & No & \multirow{2}{*}{Design} & \multicolumn{3}{c|}{Overheads*} 
& \multicolumn{3}{c|}{Security Evaluation*} & \multirow{2}{*}{Technique} \\
 & 
Agnostic & Agnostic & Re-compile & & Area & Power & Performance  &  \Basic & \Educated & \Advanced & \\ 
\midrule
WDDL~\cite{hwang2006aes}                            
&  &  & \checkmark & VLSI & 200\%      &    300\%   & 300\%      & $128\times$ & -  & -  & Power Balancing    \\
\hline
IVR~\cite{kar2017improved}                       
&  & \checkmark     & \checkmark      & VLSI & 100\%         & 100\%          & 0\%                                                & 
$\ddagger$ %
& - & - & Voltage Regulation  \\
\hline
False-Key~\cite{falsekey}                       
&  & & \checkmark & VLSI & 3\%           & 0\%      & 2\%       & $187\times$  & -  & - & Gate-Level Masking  \\
\hline
ASNI~\cite{ansi}                           
& & \checkmark & \checkmark  & VLSI & 60\%          & 68\%           & 0\%       & $1000\times$ & $1000\times$  & - & Noise Injection  \\
\hline
Blinking~\cite{althoff2018hiding}          
& \checkmark & \checkmark & \checkmark & VLSI/SW &  -           &  -               & 270\%     & $10-100\times$  & $10-100\times$ & - & Power Hiding    \\
\hline
PARAM~\cite{param}              
& \checkmark  & \checkmark & \checkmark  & µarch & $\sim$20\%          &  -              & -     & 
$\ddagger$ %
& -  & -  & Data Obfuscation \\
\hline
ARDPE~\cite{ge2020power}                
& & \checkmark & \checkmark & µarch & 7.23\%          &  -               & 3.4\%     & $4000\times$ & -  & -    & Data Randomization\\
\hline
RIJID~\cite{rijid}                  
& \checkmark  & \checkmark &   & µarch/SW & 2\%           & 27\%           & 30\%       & $\ddagger$ & -  & - & Random Code Injection \\
\hline
Block Shuffler~\cite{shuffler} 
& \checkmark & \checkmark &  & µarch/SW & 2\% & 1.5\% & 0.7\% & $\ddagger$ & - & - & Coarse Instr. Shuffling\\
\midrule
\midrule
random-iso-perf 
& \checkmark & \checkmark & \checkmark& µarch &  $\sim$0\%& $0.8\%$  & 3.8\%& \isoperb{} & \isopere{} & \isopera{} & Fine Instr. Re-ordering \\
\hline
random-iso-security
& \checkmark & \checkmark  & \checkmark & µarch & $\sim$0\%& $0.4\%$ & 11\%& - & - & \isoseca{} & Fine Instr. Re-ordering \\
\hline
random-aggressive 
& \checkmark & \checkmark  & \checkmark & µarch &  $\sim$0\% & $0\%$ & 29\% & \aggb{}  & \agge{} & \agga{} & Fine Instr. Re-ordering \\
\hline
\name{} (this work)
& \checkmark & \checkmark & \checkmark & µarch & 0.7\%     & 1.1\%            & 3.7\%                   & \parab{}  & \parae{} &  \paraa{} & Fine Instr. Re-ordering \\
\bottomrule        
\end{tabular}} %
\label{tb:ohc_pluseval}
\end{table*}

\subsection{Power and Area Overheads}
Figure~\ref{fig:power_area} shows the power and area of \name{} compared to the ooo-baseline. The Slack Unit and the GaloisLFSRs (for randomizing the delay) introduces a negligible overhead of 1.1$\%$ on \name{} over the baseline core. In matters of area, the total overhead of \name{} compared to the ooo-baseline is 0.7$\%$ that comes from the Slack Unit, the and GaloisLFSRs, the delay controller on every issue slot and the wire connections for all the new components.

\section{Related Work}
\label{sec:related}

Table~\ref{tb:ohc_pluseval} summarizes state-of-the-art countermeasures, and compares their reported performance, power, area, security evaluation and other implementation-based features. %

\subsection{Power Analysis Attack Countermeasures}

\textbf{Obfuscated execution}. One solution to overcome power analysis attacks is to obfuscate the execution and operate on the data in an obfuscated form~\cite{non-determ}. A solution for encryption engines implemented on a CGRA is proposed in ARDPE~\cite{ge2020power}. To make power and data correlation more difficult between different encryption rounds, the register usage of instructions and the data going to the encryption engine is randomized. %
Although ARDPE is one of the lowest overhead techniques (less than 10\% performance and area overhead), it specifically targets encryption engines that take plaintexts as input. Compiler support and more hardware structures will be required to apply this technique for general-purpose processors. %
More generic solutions have been proposed with~\cite{althoff2018hiding, param} aiming to secure general-purpose RISC-V in-order cores. In~\cite{althoff2018hiding}, the authors propose a hardware-software co-operative technique to detect the leakiest moments during encryption, and using a software controller to electrically disconnect the core from the system. During the disconnection, the power traces will show no correlation with the data that is being processed, however this technique can increase execution time by as much as 2.7$\times$.
PARAM~\cite{param} is another solution that investigates the leakiest modules of a RISC-V design and addresses them separately with appropriate countermeasures. 
A part of the leakages they work on are the translations in EDA tools (like Bluespec compiler~\cite{nikhil2004bluespec}), hence they modify the RTL code to prevent these leakages.
They also address the leakages of the data in the register file and buffers by obfuscating and de-obfuscating the data upon each access. In other words, they always keep the data in the obfuscated form and de-obfuscate it when they need to process the data.
The authors do not report the performance impacts of their countermeasures for the leakage they address.

\textbf{Random code injection}. In RIJID~\cite{rijid}, the authors propose a hardware-software co-design that inserts random (irrelevant) instructions in the execution at random intervals. They use the compiler to detect the regions of the code that need protection, and the hardware injects random instructions during the execution of the specified regions. Inserting real instructions in the processor during normal execution results in increased execution times. Specifically in RIJID, the performance overhead is 30\% over their non-secure baseline. Others have adopted similar approaches with code injection with similar outcomes~\cite{code-inject, random-code-inject}. 

\textbf{Random instruction shuffling}. In Block Shuffler~\cite{shuffler}, the authors propose shuffling independent blocks of instructions in a random order. In their hardware-software co-operative design, the compiler detects the independent blocks of the code and inserts \texttt{shuffle} instructions that allow the \textit{shuffler unit} in an in-order processor to generate a random permutation of the order of instruction blocks and start fetching instructions based on that order. The reported overhead in performance, power and area is negligible. The performance overhead is low because different rounds of an encryption algorithm can execute independently in any order without any impact on the control flow and data flow. The processor only requires the knowledge from the compiler.
The granularity of the shuffling in this technique is an instruction block. In \name{}, we shuffle the instructions in a finer grained way and take advantage of the slack for each instruction.
These two techniques are orthogonal and can be combined. However, the block shuffling technique requires re-compilation and modifications in the code in order to inform the processor which instruction blocks are independent and can be re-ordered.
Also, this technique might not be as effective in out-of-order cores, since these cores re-order the execution of instructions for performance reasons and could undo parts of the effects of block shuffling, especially when the instruction blocks are much smaller than the instruction window. There are also hardware- and algorithm-specific shuffling techniques to overcome power analysis attacks \cite{sac}.

\textbf{Circuit-level protection}. Power balancing techniques try to minimize the side-channel leakage by balancing the power consumption of the core at all times of the execution~\cite{bucci2006dualrail,sokolov2005design, nassar2010bcdl, tiri2002dynamic, hwang2006aes}. Previous work implements \textit{gate-level masking} that uses complex logic gates~\cite{reparaz2015consolidating, popp2007evaluation}. However, existing power balancing and masking techniques are not generic for all encryption algorithms, and they have high overheads for performance, power, and area. Other works~\cite{falsekey} use a lightweight technique that combines power balancing and hardware masking is proposed. However, it is specific for AES engines that use a fixed key. Noise injection and power isolation are other techniques that hide the intermediate values of encryption operations~\cite{guneysu2011generic, tokunaga2009secure, kar2017improved, kar2016integrated, ansi, ppe}. The authors of ANSI~\cite{ansi} combine these two techniques to implement a generic solution for encryption algorithms with negligible performance overhead. 
In summary, each of the generic solutions evaluated in this work demonstrate high power and/or area overheads. 
In addition they are designed for encryption-specific hardware only.

\subsection{Security Evaluation}

As discussed in Section~\ref{sec:security}, we consider three types of adversaries for security evaluation: (1) a \basic{} evaluation where the adversary is given a completely wrong model which gives a false sense of security, (2) an \educated{} evaluation where the given security model fits the type of countermeasure, and (3) an \advanced{} evaluation where the adversary extracts the actual leakage model. Our investigations show that none of the papers in Table~\ref{tb:ohc_pluseval} perform the \advanced{} (profiling/multivariate/worst-case) analysis and most of the reported analyses fall into the \basic{} category.

WDDL~\cite{hwang2006aes} evaluates the effect of their technique using basic CPA with Hamming model (HW for the unprotected and HD model for protected devices) and claim 128$\times$ security increase. However, using the Hamming model is inherently flawed for power-balancing-like countermeasures, as now the leakage does not correspond to the bit switch value anymore~\cite{bhasin2010countering}. For this countermeasure, the adversary has to model the function $f(x)$ that shows the consumption difference of $x-\bar{x}$. This can be done using non-profiled linear regression~\cite{schindler2005stochastic}, or profiled attacks~\cite{chari2002template}. As a result, the evaluation of this paper falls into the \basic{} evaluation. The False-Key~\cite{falsekey} paper also suffers from the same issue as the WDDL paper, since they use a Hamming model. Their analysis is performed using basic CPA in two stages: (1) targeting the false key at the Sbox output, and (2) targeting the actual key during the re-computation phase which uses power balancing. The security benefits are only brought by the second phase. Moreover, they claim their countermeasure is a masking technique, but all the leakages are first-order from a statistical point of view. They do not take into account that an advanced adversary can reverse-engineer the fixed pre-computed Sboxes and recover the secret key this way. Hence, the security evaluation in this paper falls into the \basic{} category.

The security evaluation of the IVR~\cite{kar2017improved} paper also falls into the \basic{} category. They use standard CPA and T-tests~\cite{tvla} and report a 20$\times$ security increase. Since their countermeasure adds jitter, using basic univariate methods cannot be very effective in recovering the secret key and it can be improved by using methods like simple averaging. Moreover, the tests on the protected implementation are negative (inconclusive) which do not allow one to properly quantify the security improvement. For example, rank estimation methods~\cite{poussier2016simple} could have been used to provide an improved security trend.

In PARAM~\cite{param}, basic CPA is used for security evaluation. They report that they break the secret key within 60K measurements, while the protected implementation is resistant with up to 1 million measurements. Due to the nature of their countermeasure, evaluating the security with basic CPA is good to see the effect of the presence of non-obfuscated leakage. However, as explained for IVR, negative (inconclusive) results are not sufficient to conclude security. Moreover, the security of the obfuscation itself is not investigated. Therefore, we classify the evaluation of PARAM as \basic{}. ARDPE~\cite{ge2020power} paper also uses basic CPA for evaluation and it falls into the \basic{} evaluation category. They show 4000$\times$ security increase compared to the unprotected implementation. However, their countermeasure adds desynchronization and using basic CPA is a weak evaluation method which provides a false sense of security, since an advanced adversary can reveal the key with fewer measurements.

The security evaluations of ANSI~\cite{ansi} and Blinking~\cite{althoff2018hiding} papers fall into the \educated{} category. ANSI uses basic CPA and reports 1000$\times$ security increase. Since the effect of their countermeasure is basically an SNR reduction, without any changes on the power model, evaluating the security using basic CPA is fairly adequate. The Blinking paper uses the T-test and the sum of mutual information over all samples, which shows a security increase of $10\times$-$ 100\times$ for different implementations. They are also reducing SNR without any changes in the power model. Hence, their security evaluation is considered to be the \educated{} model, just as the ANSI paper. The authors of the RIJID~\cite{rijid} and Block Shuffler~\cite{shuffler} papers do not mount actual attacks to recover the secret keys. For example in Block Shuffler~\cite{shuffler}, 100K traces are used to attack the protected core with basic DPA attack and they are not able to recover the key. But the authors do not provide the actual number of traces required for a successful attack.

\section{Conclusion}
\label{sec:conclusions}

In this work we take on the challenge of designing an efficient, secure and general-purpose processor that can protect all executing applications against side channel attacks, without affecting their performance. We propose a secure, fine-grained scheduling algorithm that dynamically reorders non-critical instructions in a random but calculated manner to desynchronize the execution and create non-deterministic behavior that increases the measurement noise. To achieve this, we exploit the time between operand availability of critical instructions (\textit{slack}) to create high-performance random schedules. In addition, we provide a comprehensive security evaluation model that includes three security evaluation standards to demonstrate more robust attacks. Our proposed solution, \name{}, offers a stronger security guarantee than demonstrated in previous works, even when tested on a more advanced and realistic security evaluation that complies with the highest security standards. \name{} improves security against power analysis attacks by \paraa{} to \parab{} with power and area overheads of 1.1\% and 0.7\% respectively. Moreover, our system achieves performance within 96\%, on average, of the baseline unprotected processor.

\bibliographystyle{plain}
\bibliography{refs}

\end{document}